\documentclass[12pt]{article}
\usepackage{amsmath,amsfonts, epsf,epsfig,color,latexsym,cite}
\usepackage{bbold}
\numberwithin{equation}{section}
\newcommand{\1}{\mathbb{1}}
\def\be{\begin{equation}}       \def\eq{\begin{equation}}
\def\ee{\label{abc}  \end{equation}}         \def\eqe{\label{abc}  \end{equation}}

\def\bea{\begin{eqnarray}}      \def\eqa{\begin{eqnarray}}
\def\ena{\end{eqnarray}}        \def\eea{\end{eqnarray}}
                                \def\eqae{\end{eqnarray}}

\def\a{\alpha}
\def\b{\beta}

\def\d{\delta}
\def\g{\gamma}

\def\p{\pi}                % Also, \varpi
\def\s{\sigma}                                   %     \varsigma

\def\G{\Gamma}

\def\P{\Pi}

\def\ca{{\cal A}}

\def\cg{{\cal G}}

\def\ci{{\cal I}}
\def\cj{{\cal J}}

\def\cl{{\cal L}}
\def\cm{{\cal H}}

\def\cp{{\cal P}}

\def\car{{\cal R}}
\def\cs{{\cal S}}

\def\cv{{\cal V}}

\def\bop#1{\setbox0=\hbox{$#1M$}\mkern1.5mu
        \vbox{\hrule height0pt depth.04\ht0
        \hbox{\vrule width.04\ht0 height.9\ht0 \kern.9\ht0
        \vrule width.04\ht0}\hrule height.04\ht0}\mkern1.5mu}
\def\pa{\partial}                              % curly d
\def\we{\wedge}                                         % wedge
\def\>{\rangle} %right angle

\def\<{\langle} %left angle

\def\Tilde#1{\widetilde{#1}}                   % big tilde
\def\to{\rightarrow}

\def\pa{\partial}

\def\half{{1 \over 2}}

\def\ha{\frac12}                               % 1/2

\def\IZ{\relax\ifmmode\mathchoice
{\hbox{\cmss Z\kern-.4em Z}}{\hbox{\cmss Z\kern-.4em Z}}
{\lower.9pt\hbox{\cmsss Z\kern-.4em Z}} {\lower1.2pt\hbox{\cmsss
Z\kern-.4em Z}}\else{\cmss Z\kern-.4em }\fi}
\def\IC{\relax\hbox{$\inbar\kern-.3em{\rm C}$}}
\def\IR{\relax{\rm I\kern-.18em R}}

\ifx\ltimes\undefined
\newcommand{\ltimes}{{\kern3pt\hbox{\vrule width 0.4pt height 5.30pt
depth .0pt}\kern-1.76pt\times\kern1pt}} \fi

\def\be{\begin{equation}}
\def\ee{\label{abc}  \end{equation}}
\def\ba{\begin{eqnarray}}
\def\ea{\end{eqnarray}}
\def\bq{\begin{quote}}
\def\eq{\end{quote}}

\def\part{\partial}

\def\beq{\begin{equation}}
\def\eeq{\label{abc}  \end{equation}}
\def\beqa{\begin{eqnarray}}
\def\eeqa{\end{eqnarray}}

\def\Ti{\tilde}
\def\ti{\Tilde}

\def\Z {\mathbb{Z}}
\def\R {\mathbb{R}}

\def\XX {\mathbb{X}}

\def\cx{{\XX}}

\begin{document}
\thispagestyle{empty}
\begin{flushright}
hep-th/0406102\\
Imperial/TP/3-04/13
\end{flushright}\vskip 0.8cm\begin{center}
\LARGE{\bf   A Geometry for Non-Geometric String Backgrounds}
\end{center}
\vskip 1in

%\vskip 0.8cm
\begin{center}{\large C M  Hull }
\vskip 0.2cm{ Theoretical Physics Group,  Blackett  Laboratory, \\
Imperial College,\\ London SW7 2BZ, U.K.}

\end{center}
\vskip 1.0cm

\begin{abstract}\noindent
A geometric string solution has background fields in overlapping
coordinate patches related by diffeomorphisms and gauge
transformations, while for a non-geometric background this is
generalised to allow transition functions   involving duality
transformations. Non-geometric string backgrounds arise from
T-duals and mirrors of flux compactifications, from reductions
with duality twists and from asymmetric orbifolds. Strings in \lq
T-fold' backgrounds with a local    $n$-torus fibration  and
T-duality transition functions in $O(n,n;\Z)$ are formulated in an
enlarged space with a $T^{2n}$ fibration
 which is geometric, with spacetime emerging locally from a
choice of a $T^n$ submanifold of each $T^{2n}$ fibre, so that it
is a subspace or brane  embedded in the enlarged space. T-duality
acts by changing to a different $T^n$ subspace
of $T^{2n}$. For a geometric background, the local
choices of $T^n$ fit together to give a spacetime which is a $T^n$
bundle, while for non-geometric string backgrounds they
do not fit together to form a manifold. In such cases spacetime
geometry only makes sense locally, and the global structure
involves the doubled geometry. For open strings, generalised
D-branes wrap a $T^n$ subspace of each $T^{2n}$ fibre and the
physical D-brane is the part of the physical space
lying in the generalised D-brane subspace.
\end{abstract}

\vfill

\setcounter{footnote}{0}
\def\thefootnote{\arabic{footnote}}
\newpage

\section{Introduction}\label{Introduction}

There is   evidence that string theory can be consistently defined
in non-geometric backgrounds in which the transition functions
between coordinate patches involve not only diffeomorphisms and gauge
transformations but also duality transformations [1-7].
%\cite{Dabholkar:2002sy}-\cite{Fidanza:2003zi}.
Such backgrounds can arise from compactifications with duality twists
\cite{Dabholkar:2002sy},\cite{Flournoy:2004vn},\cite{Hull:1998vy},\cite{Hull:2003kr} or
from acting on   geometric backgrounds with fluxes with
  T-duality \cite{Kachru:2002sk},\cite{Hellerman:2002ax},\cite{Lowe:2003qy}
  or  mirror symmetry \cite{Gurrieri:2002wz},\cite{Fidanza:2003zi}.
In special cases, the compactifications with duality twists are
equivalent to asymmetric orbifolds \cite{Dabholkar:2002sy} which
can give consistent string backgrounds
\cite{Narain:1986qm},\cite{Narain:1990mw},\cite{Aoki:2004sm}. It
is to be expected that transition functions involving duality
transformations should arise since dualities are fundamental
symmetries of the theory on the same footing as diffeomorphisms
and gauge transformations. Not all such backgrounds are
consistent in the quantum theory, and they must be restricted by demanding conformal,
Lorentz and modular invariance on the world-sheet.
 Understanding how conventional
geometry must be generalised to incorporate such backgrounds could
provide important clues as to the nature of string theory or
M-theory, as they require going beyond the familiar formulations
of field theory in a geometric spacetime or a sigma-model with
geometric target space.

Duality transformations   sometimes take a geometric string
background to a non-geometric one,  as they can lead to transition
functions between coordinate patches which are not the coordinate
transformations of standard differential geometry but are exotic
ones that involve dualities. One could take the point of view that
in such circumstances there is an obstruction to T-dualizing and
the background does not have a T-dual. However, the non-geometric
configuration that follows from formally applying the T-duality
rules seems to make sense as a string background. Locally, there
is a conventional spacetime interpretation, and the problem is in
the transition functions relating overlapping geometric
neighbourhoods. A perturbative string background is specified by a
conformal field theory (CFT), and a given CFT can arise from a
number of different sigma-models
 with very different target space geometries.
 The different target space geometries can be thought of as different representations of the
 same CFT, and so it seems reasonable to consider situations
 with transition functions that  relate different representations of the same CFT.
 In particular, T-duality changes the target space geometry
 to another giving the same CFT, and so T-duality transition
 functions should be allowed. Some examples (such as those
  related to asymmetric orbifolds) are known to give consistent backgrounds, and compactifying on a T-fold (as in
 \cite{Dabholkar:2002sy},\cite{Flournoy:2004vn}) gives
  a conventional low-energy field theory.
In this paper, it will be assumed that T-duality can be
 applied in such cases, and the properties  of the resulting non-geometric backgrounds will be explored.

 A geometric background $(M,G,B,\Phi)$ is a spacetime manifold
$M$ with metric $G$, 2-form gauge field $B$ and scalar field
$\Phi$ (possibly with other fields such as RR gauge fields) so that   the transition
functions in the overlaps between coordinate patches are
diffeomorphisms,  together with a gauge transformation so that
$B'=B+d\Lambda $, i.e. $G,\Phi, H=dB$ are tensor fields and $B$ is a gerbe
connection on $M$. If $M$ is a torus bundle with fibre $T^n$, then
each transition function  includes an element of
$GL(n,\Z)=SL(n,\Z)\times \Z_2$, the group of large diffeomorphisms
on $T^n$. However, string theory on such a background has a
T-duality symmetry $O(n,n;\Z)$, and acting on the background with
a T-duality transformation $g\in O(n,n;\Z)$ will take a transition function $S
\in GL(n,\Z)$ to $S'=gSg^{-1} $ which will in general be   in
$O(n,n;\Z)$ and need not lie in the   subgroup $GL(n,\Z)$ that acts through torus diffeomorphisms.
This  means that the new transition functions will in general mix
momentum and winding modes, so that a geometrical configuration in
one patch can translate into a stringy one featuring winding modes
in the next. T-duality then indicates that transition functions in
$O(n,n;\Z)$ should be allowed, and such configurations in general
are not geometric backgrounds. The purpose
here is to explore such non-geometrical backgrounds and present a
formulation that captures the novel structure and allows T-duality
to act.
The full transition functions for the torus bundles that will be considered here  are in $GL(n,\Z)\ltimes U(1)^{n}$, where each $U(1) $ acts as a translation on a circle fibre.

This notion of what is meant by a geometric background could be broadened to allow  transition functions that include integral shifts of the $B$-field, $B \to B+ \Theta$ where $\Theta \in H^2(T^n,\Z)$, so that $B$ is a gerb-connection.
These shifts are part of $O(n,n;\Z)$ and so such transition functions are included in the class of backgrounds considered here, and
so one could choose to define the subset of backgrounds with classical geometries as either those with diffeomorphism transition functions, or to include those with $B$-twists.
 It will be convenient here  to define the \lq geometric subgroup' of $O(n,n;\Z)$ to be
$GL(n,\Z)$ rather than the group  $GL(n,\Z)\ltimes \Z^{n(n-1)/2}$ with $B$-shifts.

It is convenient to combine the metric $G_{ij}$ and  2-form $B_{ij}$
on an $n$-torus
 into the $n\times n $ matrix $E_{ij}=G_{ij}+B_{ij}$. These moduli will be taken to be constant
 on $T^n$ but allowed to depend on the coordinates of the base space $N$.
A transition function in $ GL(n,\Z)$   represented by an $n\times n $ matrix $h_i{}^j$
will relate the values of  $E$ in two patches in $N$  by
\begin{equation}
E'=hEh^t
\label{abc}  \end{equation}
so that $G'=hGh^t$, $B'=hBh^t$.
Consider now an element of $O(n,n;\Z)$
\begin{equation}
g=\left(
\begin{array}{cc}
a & b\\
c & d \end{array}\right),
\label{abc}  \end{equation}
where $a,b,c,d$ are integer-valued $n\times n$ matrices, and such that
$g$ preserves the metric $L$
\begin{equation}
L=\left(\begin{array}{cc}
0 & \1  \\
\1 & 0
\end{array}\right),
\label{dsfsdh}
 \end{equation}
The transformation of $E$ under a T-duality transformation $g\in O(n,n;\Z)$ is non-linear
\cite{Buscher:sk},\cite{rocek},\cite{Giveon:1994fu}:
\begin{equation}
E'  = (aE+b)(cE+d)^{-1}.
\label{Etrans}  \end{equation}
while the dilaton transforms as
\begin{equation}
e^{\Phi '}=e^{\Phi }{\left( \frac {{\rm det} ~G' } {{\rm  det}
~G}\right) }^{1/4} \label{fitrans}  \end{equation} A transition
function   in $O(n,n;\Z)$ then relates $(E,\Phi)$  and
$(E',\Phi')$ in the overlap of two patches by the non-linear
transformations (\ref{Etrans}),(\ref{fitrans}). The non-linear
transformations mix $G$ and $B$, and mix momentum modes on the
torus with string winding modes, so that a geometric description
in one patch can be translated to a non-geometric one in an
adjacent patch.

Examples arise when   there is a  T-duality monodromy around a
non-contractible loop. String theory compactified on $T^n$ has
$O(n,n;\Z)$ symmetry, and then compactification on  a further
circle can be twisted by an $O(n,n;\Z)$ monodromy, giving a
stringy generalisation \cite{Dabholkar:2002sy},\cite{Hull:1998vy}
of a Scherk-Schwarz reduction
%\cite{Scherk:1979zr}-\cite{Bergshoeff:2002mb}.
[16-22].
  If the monodromy is in the geometric
$GL(n;\Z)$ subgroup that acts through $T^n$ diffeomorphisms, then
this can be lifted  to a higher dimensional theory compactified
on a $T^n$ bundle over a circle, but for general $O(n,n;\Z)$
monodromy there is no such geometric interpretation. Locally there
is  a $T^n$ fibration  with $G,B$ defined on an internal $T^n$,
but there is no globally defined geometry on the bundle. If the
circle coordinate is  $Y$ with the identification $Y \sim Y+2\pi $,
then the  torus moduli $G,B$ and dilaton satisfy the boundary
conditions
\begin{eqnarray}
E ( Y +2\pi ) &=& \frac{  a E(Y) + b }{  cE(Y) + d } \nonumber\\
e^{\phi( Y+ 2\pi )} & = &
e^{\phi ( Y) } \left( \frac {{\rm det} ~G(Y
+ 2\pi ) }{{\rm  det} ~G( Y)} \right)^{1/4} \label{O22trans}
\end{eqnarray}
A momentum mode at $Y$ is identified with a linear combination of
momentum and  winding modes at $Y+2\pi$, so that the background is
intrinsically stringy. A large class of such twisted reductions
are equivalent to asymmetric orbifolds, and so are consistent
string theory backgrounds. Modular invariance imposes an important
constraint on such backgrounds, but those that arise as duality
transforms of geometric backgrounds will be modular invariant.

Such twisted reductions arise as T-duals of flux
compactifications. Consider for example reduction on a 3-torus
with constant NS $H$-flux given by an integer $m$ times the volume
form\footnote{This is not a conformal field theory as it stands. A
consistent string background is obtained by allowing the moduli to
depend on an extra coordinate, to obtain a space with topology
$T^3\times \R$. The product of this with 6-dimensional Minkowski
space can be viewed as a NS 5-brane with transverse space
$T^3\times \R$ that is smeared over the $T^3$ so that the harmonic
function in the ansatz is a function on $\R$ \cite{Hull:1998vy}.
The discussion given here should be applied to such a conformal
field theory. }. A T-duality on any circle gives a twisted
reduction on a $T^2$ bundle over a circle, with geometric
monodromy  
\begin{equation}
  \begin{pmatrix} 1&m \\ 0&1  \end{pmatrix}
\label{abc}  \end{equation} in $GL(2,\Z)$. Then a   T-duality on
one of the $T^2$ directions takes one back to the $T^3$ with flux
\cite{Hull:1998vy}, while the T-duality on the other fibre
direction gives a background with monodromy in $O(2,2,\Z)$ given
by \cite{Kachru:2002sk},\cite{Lowe:2003qy}
\begin{eqnarray}
 \begin{pmatrix}1 & 0 & 0 & 0 \\
                      0 & 0 & 0 & 1 \\
                      0 & 0 & 1 & 0 \\
                      0 & 1 & 0 & 0   \end{pmatrix}
\end{eqnarray}
so that the transition functions are of the form (\ref{O22trans}).

As mirror symmetry is related to a T-duality on the fibres of
Calabi-Yau spaces with a $T^3$ fibration \cite{Strominger:1996it},
the mirror of a Calabi-Yau with NS flux on its $T^3$ fibres will
in general be non-geometric of this type \cite{Fidanza:2003zi}.
Similar examples can also arise with cosmic strings in 4
dimensions or $p$-branes in $D=p+3$ dimensions with a duality
monodromy around the string or brane. The general feature in all
such examples are transition functions involving a duality
symmetry. These duality symmetries could be T-dualities or
U-dualities (e.g. in spaces   with torus fibrations) or mirror
symmetries (in spaces   with Calabi-Yau fibrations). Backgrounds
with T-duality transition functions will be referred to as
T-folds, while those with S-duality, U-duality or mirror symmetry
transition functions will be referred to as S-folds, U-folds or
mirror-folds etc. Examples of S-fold compactification were
investigated in \cite{Hull:2003kr}. In this paper, attention will
be restricted to the case of  T-folds, as these provide
backgrounds that can be analysed  in perturbative string theory.

For string theory compactified on an $n$-torus $T^n$, the internal
momentum $p^i$ combines naturally with the winding number $w^i$ to
form the momenta $p_L=p+w$, $p_R=p-w$ taking values in the Narain
lattice. Conjugate coordinates $X^i_L$, $X^i_R$ are needed e.g. to
write vertex operators $e^{kX_L}$, $e^{kX_R}$. These are given in
terms of the torus coordinates $X^i$ and the coordinates $\ti X^i$
on a dual torus by $X_L=X+\ti X$, $X_R=X-\ti X$. For each $i$, if
$X^i$ is a coordinate on a circle of radius $R_i$, then  $\ti X^i$
is the coordinate on the T-dual circle of radius $1/R_i$. The
theory is then naturally formulated as a string theory with target
space the doubled torus $T^{2n}$ with coordinates $X,\ti X$, and
the T-duality symmetry $O(n,n;\Z)$ acts naturally on $T^{2n}$.
This doubles the degrees of freedom, but they can be halved again
by imposing the constraints that $X_L$ is left-moving and $X_R$ is
right moving. This can be written in terms of the pull-back
world-sheet one-forms $dX=\pa _\a X  d\s ^\a $ as the self-duality
conditions
\begin{equation}
dX_L= *dX_L, \qquad dX_R= -*dX_R
\label{LRcons}  \end{equation}
where $*$ is the world-sheet Hodge dual.

The dual coordinates $\ti X$ are needed also for string field
theory, as the string field $\Psi[X(\s)]$ for strings on a torus
is a field $\Psi[X , \ti X, a,\Ti a]$ depending on the coordinates
$X^i,\ti X^i$ as well as the Fourier coefficents $a_n^i, \Ti
a_n^i$ \cite{Kugo:1992md}. This is usually expanded in terms of an
infinite set of fields $\psi(X)$ on $T^n$, but could instead be
expanded in terms of an infinite set of fields $\psi(X,\ti X)$ on
the doubled torus $T^{2n}$, and  if there are transition functions
that mix $X,\ti X$, then this latter formulation seems the more
natural.

This suggests using a doubled formalism with a doubled torus
$T^{2n}$ containing both the original torus with coordinates $X^i$
and a dual $n$-torus with coordinates $\ti X_i$. There is a
natural $O(n,n)$ invariant metric $ds^2_L=2dX^id\ti X_i$ on this
space and the subgroup of the   group $GL(2n,\Z)$ of  large
diffeomorphisms of $T^{2n}$ that preserve this metric is
$O(n,n;\Z)$. Then $O(n,n;\Z)$ transition functions between patches
in the base space $N$ are part of the data needed to construct a $T^{2n}$
bundle over $N$ with natural fibre metric  $ds^2_L$. Thus the
natural geometry that can be constructed from T-fold data is one
in which the $T^n$ bundle over $N$ is replaced by a $T^{2n}$
bundle over $N$ which, as will be seen, can be viewed as a \lq
universal space' containing all possible T-duals of the original
configuration. T-duality transition functions that glue momentum
modes in one patch to linear combinations of momentum and winding
modes in another patch then become geometric for the enlarged
space in which the winding modes are represented geometrically on
the dual $n$-torus, as the transition functions are now
diffeomorphisms of $T^{2n}$. 
The transition functions for such a $T^{2n}$ bundle are in 
$O(n,n,\Z)\ltimes U(1)^{2n}$.
This doubling of course increases the
degrees of freedom, and the right counting is obtained by imposing
a covariant self-duality constraint that generalises
(\ref{LRcons}) and which implies, roughly speaking, that half of
the coordinates on the $T^{2n}$ correspond to right-movers and
half to left-movers, although the split into left and right moving
chiral bosons depends on the position in $N$. The doubled
formalism involving both the $X$ and the dual coordinates $\ti X$
is essentially that of Cremmer, Julia, Liu and Pope
\cite{Cremmer:1997ct}, and related to those of \cite{Maharana:1992my},
\cite{Duff:1990}. If the chiral boson constraint is
implemented using the non-covariant formalism of
\cite{Floreanini:1987as}, then the  formulation presented here is
closely related to that proposed by Tseytlin
\cite{Tseytlin:1990nb},\cite{Tseytlin:1990va} and used in
\cite{Schwarz:mg}, while the approach to chiral bosons of
 \cite{Siegel:1983es}  leads to a formulation similar to that of \cite{Hull:si}.

The physical space is then a subspace $T^n \subset T^{2n}$, which
might be thought of as a kind of $n$-brane in $T^{2n}$, and  T-duality
can be understood as changing which $T^n$ subspace is the physical
space. The $T^n$ is a completely null subspace with respect to the
$O(n,n)$ metric i.e. all tangent vectors are null.
 Locally, over each coordinate patch $U$ in
$N$ one can choose a local $T^{n}$ slice of the $T^{2n}$ bundle,
and a conventional geometric picture arises on the local patch of spacetime which
is a (trivial) $T^n$ bundle over $U$. If these can be patched
together to form a $T^n$ bundle over $N$, then there is a
geometric string background, but in general there will not be a
global $T^{n}$ slice of the $T^{2n}$ bundle, and there is no way
of picking out a spacetime that is a $T^n$ bundle over $N$. Then
the local spacetime patches (a trivial $T^n$ bundle over $U$ for each
patch $U \subset N$) do not patch together to form a manifold,
but instead form a T-fold.

The plan of the paper is as follows. In the next section the
doubled formalism is introduced in which the torus fibres are
doubled from a $T^n$ to a $T^{2n}$ and string theory on a space
which is locally a $T^n$ bundle over each patch $U$ in $N$ is rewritten as a string
theory on a space $\ti M$ which is a $T^{2n}$ bundle over $N$
subject to a self-duality constraint. This formulation is
manifestly invariant under the T-duality group $O(n,n;\Z)$.
Choosing a polarization, i.e. a local choice of $T^n\subset
T^{2n}$ leads to the conventional formulation and to the standard
Buscher T-duality transformations of the sigma-model geometry. In
section 3, the $n=1$ case is analysed and developed as an explicit
example of the general constructions of later sections. In section
4, the action of T-duality is analysed and it is shown that it
changes the polarization, i.e. it transforms the physical $T^n$
subspace of $T^{2n}$ to a different subspace, so that T-duality is
the statement that the physics is independent of the choice of
polarization. In section 5, polarizations are characterised
geometrically in terms of product structures. In section 6, the topology of T-folds and their
doubled formulation is carefully analysed, and it is seen that for
a geometric background there is a global polarization, i.e. there
is a $T^n$ bundle over $N$ arising as a subspace of the doubled
space  $\ti M$, and this subspace is the geometric spacetime.
However, for non-geometric backgrounds, one can choose a
polarization locally but the  physical patches of spacetime do not patch together to form a
manifold, so that there is a local notion of spacetime, but not a
global one. In section 7, the formalism is extended to open
strings and D-branes. As T-duality can take a D$p$-brane to a
D$p'$-brane with $p\ne p'$, T-duality transition functions can
glue a D$p$-brane to a D$p'$-brane, and it is shown how this
naturally emerges from the formalism.
 In the final section, a number of further issues are discussed, including quantization, the
 generalisation to U-duality   and the relationship with
 F-theory.

\section{The Duality-Invariant Action}
\subsection{The Doubled Formalism}

In this section  a formalism is  presented for sigma-models whose
target space $M$ is locally a $T^n$ bundle over a base space $N$.
 The $T^n$ is considered as a subspace of a doubled torus
$T^{2n}$ containing both the original torus and a dual $n$-torus
and   an enlarged  target space $\ti M$ is introduced which is a
$T^{2n}$ bundle over $N$. The discussion in this section is local,
considering the spacetime as $U\times T^n$ for some coordinate
patch $U$ in $N$, embedded in an enlarged space $U\times T^{2n}$.
The global issues involved in gluing such patches together will be
analysed in section 6.
 Let $Y^m$ be
coordinates on (a patch $U$ of) the base $N$ and $\cp ^I= \cp
^I_\a d \s ^\a$ ($I,J=1,\dots ,2n$) be the covariant momenta on
$T^{2n}$ that are world-sheet one-forms and tangent vectors on
$T^{2n}$. They satisfy  the Bianchi identity
\begin{equation}
d \cp ^I =0
\label{bian}  \end{equation}
so that locally there are
coordinates $\cx ^I$ with
\begin{equation}
  \cp ^I _\a =\pa _\a \cx ^I
\label{abc}  \end{equation} Coordinates on $T^{2n}$ will be chosen
to satisfy periodicity conditions, so that $\cx ^I \in \R ^{2n}/
\G$ for some $2n$-dimensional lattice $\G$. This choice fixes the
reparameterization invariance up to large diffeomorphisms, the
group $GL(2n,\Z)$ of automorphisms of the lattice $\Gamma$.
 The
world-sheet lagrangian is
\begin{equation}
\cl _d= \half \cm_{IJ} \,  \cp ^I \we * \cp ^J + \cp ^I \we * \cj
_I + \cl (Y) \label{lagd}  \end{equation} where $\cm_{IJ}(Y)$ is a
positive-definite metric on $T^{2n}$ that can depend on the base
coordinates $Y$, $*$ denotes the world-sheet Hodge dual and $ \cj
_I=\cj _{I\a}(Y)d\s ^\a$ are source terms that depend on $Y$ but
are independent of $\cx$, while $\cl (Y)$ is the remaining
$\cx$-independent part of the lagrangian. To simplify the
discussion, the world-sheet will be taken to be flat for now, so
that terms involving the curvature such as the Fradkin-Tseytlin
term do not appear.
As will be discussed in what follows,   this theory is    equivalent to a standard sigma-model formulation
 if the metric $\cm$ is restricted to be a natural metric on the coset $O(n,n)/O(n)\times O(n)$.
The field equation from varying $\cx$ is
\begin{equation}
d*(\cm \cp +\cj)=0
\label{feqn}  \end{equation}

The number of fibre coordinates has been doubled, so  a
self-duality constraint generalising (\ref{LRcons}) is imposed
that halves the degrees of freedom. The constraint is
 \begin{equation}
 \cp^I = L^{IJ} *(\cm _{JK}\cp^K +\cj _J)
\label{cons}  \end{equation}
 for some constant invertible symmetric matrix $L^{IJ}$.
  Then this together with  the Bianchi identity (\ref{bian}) implies the field equation (\ref{feqn}).
 Consistency of the constraint requires
  \begin{equation}
S ^2 = \1
\label{abc}  \end{equation}
where
 \begin{equation}
S^{I}{}_J= L^{IK}\cm _{KJ}
\label{abc}  \end{equation}
 and
 \begin{equation}
\cm L*\cj = - \cj
\label{jcons}  \end{equation}
Then $S=L\cm$ has eigenvalues $\pm 1$, and
we choose $L_{IJ}$ (the inverse of  $L^{IJ}$)
to be an $O(n,n)$ invariant metric.
(For heterotic  strings, the metric $L$ would be   $O(n,n+16)$ invariant.)
For a flat target space, $\cm =\1$ and the constraint reduces to (\ref{LRcons}).

The lagrangian is manifestly invariant under the  rigid  $GL(2n, \R)$
transformations
\begin{equation}
\cm \to g^t \cm g, \qquad \cp \to g^{-1} \cp, \qquad \cj \to g^t \cj
\label{gtrans}  \end{equation}
(with $Y$ and $\cl (Y)$ invariant).
The corresponding transformation of the coordinates
\begin{equation}
  \cx \to g^{-1} \cx
\label{xtrans}  \end{equation}
only preserves the boundary conditions
if $g$ is restricted to be in the subgroup $GL(2n,\Z) \subset GL(2n, \R)
$ preserving the lattice $\G$.
The constraint (\ref{cons}) breaks $GL(2n, \R)$
   to the subgroup $O(n,n)$ preserving $L^{IJ}$ and so breaks $GL(2n,\Z)$ to
   $O(n,n;\Z)$.
   Thus this formulation is manifestly invariant under the T-duality group
      $O(n,n;\Z)$.

  To understand the significance of $\cj$, it is useful to decompose it into a part proportional to $dY$ and one  
proportional to $*dY$
 \begin{equation}
 \cj _I =  \cm _{IJ}(\ca ^J+*\ti \ca^J)
\label{asdasd}  \end{equation}
where
\begin{equation}
\ca^I=\ca^I_mdY^m, \qquad  \ti \ca ^I=\ti \ca ^I_mdY^m
\end{equation}
Then the constraint (\ref{jcons})
becomes
\begin{equation}
S\ca =-\ti \ca
\end{equation}
giving $\ti A$ in terms of $A$, 
while
(\ref{cons})
becomes
\begin{equation}
(\cp +\ca)= S*(\cp +\ca)
\end{equation}
The $\ca^I_m$ are the 
components of a connection for the $T^{2n}$ fibration over $N$, whose 
pull-backs to the world-sheet is $\ca^I$, so that
$\cp +\ca$ is a covariantized momentum, and the lagrangian could be written in terms of
$\hat \cp= \cp +\ca$ instead of $\cp$ as
\begin{equation}
\cl _d= \half \cm_{IJ} \, \hat  \cp ^I \we *\hat  \cp ^J +   \cl '(Y) \label{lagdh}  \end{equation}
The connection transforms as
\begin{equation}
  \ca \to g^{-1} \ca
\label{atrans}  \end{equation}
under $O(n,n)$.

\subsection{Right and Left Movers}

As will be seen below, the doubled theory is equivalent to a
standard sigma-model formulation provided the
   metric $\cm_{IJ}$ is restricted to be a
coset metric   for $O(n,n)/O(n)\times O(n)$, so that on
identifying under the action of $O(n,n;\Z)$, the moduli space for
such metrics is $O(n,n;\Z) \backslash O(n,n)/O(n)\times O(n)$. It
is convenient to parameterize the coset space $O(n,n)/O(n)\times
O(n)$ by a $2n\times 2n$ vielbein $\cv ^A {}_I(Y)$ which is an
element of $O(n,n)$ identified under the left action of
$O(n)\times O(n)$, $\cv \sim k\cv$. The field $\cv(Y)$ then
transforms as
\begin{equation}
\cv \to k(Y) \cv g \label{abc}  \end{equation} under a rigid
$O(n,n) $ transformation $g \in O(n,n)$ and a local $O(n)\times
O(n)$ transformation $k(Y) \in O(n)\times O(n)$, so that $k^tk=\1$. The vielbein can
be used to transform between the $O(n,n)$ indices $I,J$ and the
$O(n)\times O(n)$ indices $A,B$. The metric is then
\begin{equation}
\cm = \cv ^t \cv \label{abc}  \end{equation} and is manifestly
invariant under local $O(n)\times O(n)$ transformations and
transforms under $O(n,n)$ as in (\ref{gtrans}). The indices
$A,B=1,...,2n$ transform under $O(n)\times O(n)$ and can be split
into indices $a,b=1,...,n$ and $a',b'=1,...,n$ for the two $O(n)$
factors, $A=(a,a')$, so that in a natural basis
\begin{equation}
L^{AB} = \begin{pmatrix} L^{ab}& 0 \\ 0& L^{a'b'}
  \end{pmatrix} =
   \begin{pmatrix} \1 ^{ab}& 0 \\ 0& -\1 ^{a'b'}
  \end{pmatrix}, \qquad
  S^{A}{}_B = 
   \begin{pmatrix} \d ^a {}_b & 0 \\ 0& -\d ^{a'}{}_{b'}
  \end{pmatrix}
\label{abc}  \end{equation}

Then
\begin{equation}
\cv ^A {}_I =\left( \begin{array}{c} \cv  ^a {}_I \\ \cv ^{a'} {}_I \end{array}\right),
\qquad  \cv \cp =
\left( \begin{array}{c} \cp  ^a  \\ \cp ^{a'}  \end{array}\right),
\label{abc}  \end{equation}
and
\begin{equation}
\cm _{IJ}= \cv  ^a {}_I \cv  ^b {}_J \d _{ab} + \cv  ^{a'} {}_I \cv  ^{b'} {}_J \d _{a'b'}
\label{abc}  \end{equation}

The self-duality constraint can now be written as
\begin{eqnarray}
\cp ^{a} &=&+ * (\cp^{a} + \cj^{a})
\nonumber \\
\cp ^{a'} &=& -* (\cp^{a'} + \cj^{a'})
\end{eqnarray}
or
\begin{eqnarray}
\hat \cp ^{a} &=&+ * \hat \cp^{a} 
\nonumber \\
\hat \cp ^{a'} &=& -* \hat \cp^{a'} 
\end{eqnarray}

Introducing null coordinates on the world-sheet $\a =(+,-)$ so that
$\cp _\pm = \pm *\cp _\pm$, this becomes
\begin{eqnarray}
\cp ^{a}_- &=&- \ha  \cj ^{a}_- , \qquad   \cj ^{a}_+=0
\nonumber \\
\cp ^{a'}_+ &=&- \ha  \cj ^{a'}_+ , \qquad   \cj ^{a'}_-=0
\end{eqnarray}
or
\begin{eqnarray}
\hat \cp ^{a}_- &=&0
\nonumber \\
\hat \cp ^{a'}_+ &=&0
\end{eqnarray}

In the case of a trivial bundle with constant $\cv$ independent of
$Y$ and $\cj=0$,  this would imply $\pa _- X^a=0$ and $\pa _+
X^{a'}=0$ so that $X^a$ are right-movers and $X^{a'}$ are
left-movers, giving the right count of degrees of freedom. More
generally, the  $Y$ dependence   of $\cv$ means that the way the
$\cx^I$ are split into left-movers and right-movers depends on the
position $Y$ on the base, while the source $\cj$ further modifies
the constraint.

\subsection{Choice of Polarization}

The physical spacetime is locally a product of an $n$-torus with a
region of $N$, and the physical $T^n$  is embedded in   $T^{2n}$.  In
order to make contact with the conventional formulation, one needs
to choose a polarization, i.e. to choose a splitting of $T^{2n}$
into a physical $T^n$ and a dual $\ti T ^n$ for each point in $N$,
splitting the fibre coordinates into the physical coordinates
$X\in T^n$ and the dual coordinates $\ti X\in \ti T^n$, and then
write the theory in terms of the coordinates $X$ alone, solving
the constraint (\ref{cons}) to express $\ti X(\s)$ in terms of $X(\s)$.
Then the variables $X$ would be the ones integrated over in the
functional integral and invariance of the theory under T-duality
implies that the  physics is independent of the choice of
polarization.

In order to define a polarization or local product structure on
the fibres,  one first chooses a subgroup $GL(n,\R)$ of $O(n,n)$ under
which the fundamental $2n$ of $O(n,n)$ splits into  the
fundamental representation $n$ of $GL(n,\R)$  and the dual
representation $n'$, $2n \to n \oplus n'$. It will be useful to
use a superscript $i$ for the fundamental representation $n$
(where $i=1,..., n$) and a subscript $i$ for the  dual
representation $n'$, and introduce constant projectors $\P ^i
{}_I$   and    $\Tilde \P _{iI} $, so that
\begin{equation}
  \cp =\left( \begin{array}{c} \P ^i {}_I \cp ^I
    \\  \Tilde \P _{iI} \cp ^{I}  \end{array}\right)=
\left( \begin{array}{c} P  ^i  \\ Q _i  \end{array}\right).
\label{abc}  \end{equation} This can be thought of as a choice of
basis, but it is useful to introduce the projectors explicitly, so
as to keep track of the choice of subgroup  $GL(n,\R)$ of
$O(n,n)$; as we shall see, duality transformations change the
projectors and change the subgroup  $GL(n,\R)$ to a conjugate one.
The norm of $\cp$ with respect to the $O(n,n)$ metric is $\vert \cp
\vert ^2= 2 P^iQ_i$.
 The $\P ^i
{}_I$ projects the tangent space of $T^{2n}$ onto a maximally
isotropic or lagrangian subspace, i.e. a subspace which is null with respect to
the metric $L_{IJ}$ and of maximal dimension $n$, while $\Tilde \P
_{iI} $ projects onto the complementary null subspace. This is
equivalent to choosing a pure spinor for $Spin{(n,n)} $ \cite{Hitchin}.
Then
\begin{equation}
\P ^i {}_I\P ^j {}_J L^{IJ}=0,
\qquad
\Tilde\P  _{iI}\Tilde \P _{jJ} L^{IJ}=0,
\label{abc}  \end{equation}
and the metric $L$ is off-diagonal in the $GL(n)$ basis  and can be written as
\begin{equation}
L=\left(\begin{array}{cc}
0 & \1 \\
\1 & 0
\end{array}\right)
\label{abc}  \end{equation}
The $GL(n)$ subgroup is embedded in $O(n,n)$ in this basis as matrices of the form
\begin{equation}
 \left(\begin{array}{cc}
h^{-1} &0 \\
0& h^t
\end{array}\right),
\label{hing}  \end{equation}
where $h=h^i{}_j$ is an $n\times n$ matrix in $GL(n)$.

Choosing a subgroup $GL(n,\R) \subset O(n,n)$ allows a splitting
of the tangent spaces. If the projectors in fact project onto a
subgroup $GL(n,\Z) \subset O(n,n;\Z)$, then this extends to a
splitting of the coordinates $\cx^I \to (X^i, \ti X_i)$ that is
consistent with the boundary conditions, with $ P^i=dX^i$, $  Q_i=
d\ti X_i $. Then
\begin{equation}
  \cx =\left( \begin{array}{c} \P ^i {}_I \cx ^I
    \\  \Tilde \P _{iI} \cx ^{I}  \end{array}\right)=
\left( \begin{array}{c} X  ^i  \\ \ti X _i  \end{array}\right),
\label{abc}  \end{equation} with the $X^i$ the coordinates of the
$T^n$ subspace and $\ti X_i$ the coordinates of the dual $\ti T^n$
subspace.
  The $O(n,n)$-invariant
metric in these coordinates is
\begin{equation}
ds^2= 2 dX^i d\ti X_i \label{abc}  \end{equation} so that the
$T^n$ submanifold with coordinates $X^i$ is a maximally null
subspace. Choosing a polarization   splitting $T^{2n} \to
T^n\oplus \ti T^n$  then corresponds to choosing a subgroup
$GL(n,\Z) \subset O(n,n;\Z)$.

In this basis the vielbein has components
\begin{equation}
\cv ^A {}_I =
 \begin{pmatrix} \cv ^{a} {}_i& \cv ^{aj}  \\ \cv ^{a'} {}_i& \cv ^{a'j}   \end{pmatrix}
\label{abc}  \end{equation} The local $O(n)\times O(n)$ symmetry
can be used to choose a triangular gauge for $\cv$, so that
\begin{equation}
\cv= \left(\begin{array}{cc} e^t & 0 \\  -e^{-1}B
&e^{-1}\end{array}\right), \label{vist}  \end{equation} for some
$n$-bein $e_i^a$ and anti-symmetric $n\times n$ matrix $B_{ij}$.
Then
\begin{equation}
\cm= \cv ^t \cv =\left(
\begin{array}{cc}
G-BG^{-1}B & BG^{-1} \\
-G^{-1}B   & G^{-1}
\end{array}
\right).
\label{abc}  \end{equation}
where
the metric $G=e^te$, i.e.
\begin{equation}
G_{ij}= e_i{}^a e_j{}^b \d _{ab}
\label{abc}  \end{equation}
As a result, the fibre metric $\cm (Y) $ is parameterized by
an $n\times n$ matrix $E(Y)$ given by
\begin{equation}
E_{ij}=G_{ij}+B_{ij}
\label{abc}  \end{equation}
Note that the inverse of $\cm$ is $\cm^{IJ} = L^{IK}L^{JL}\cm_{KL}$ so that
\begin{equation}
\cm ^{-1}=L\cm L  =\left(
\begin{array}{cc}
G^{-1}  &  -G^{-1}B \\
BG^{-1}    & G-BG^{-1}B
\end{array}
\right).
\label{abc}  \end{equation}
and $(L\cm
)^2=\1$.

The current
$\cj $ decomposes into
\begin{equation}
J_i = \Tilde \P _{iI} L^{IJ} \cj_J, \qquad
K^i = \P ^i {}_I L^{IJ} \cj_J
  \label{abc}  \end{equation}
  so that in the $GL(n)$ basis
  \begin{equation}
  \cj   =\left( \begin{array}{c}J_i      \\ K^i  \end{array}\right)
  \label{cjist}
  \end{equation}
and
  the source-term in the lagrangian
becomes
\begin{equation}
 \cp ^I \we * \cj _I
=P^i \we *  J_i + Q_i \we * K^i \label{abc}  \end{equation}
The $X^i$ are coordinates on a torus $T^n$ and,
as we shall see, the dynamics  of $X^i$ are governed
 by a sigma-model on $T^n$ with metric $G_{ij}$
and B-field $B_{ij}$, while the $\ti X_i$ are T-dual coordinates.
If $B=0$, the metric $\cm$ restricted to on $T^n$ is $G_{ij}$
while restricted to  $\ti T^n $ it is the inverse of this,
$G^{ij}$, so that $\ti T ^n$ is the dual torus to $T^n$ with
respect to the metric $\cm$. However, the $\cv $ defined in
(\ref{vist}) can be written as
\begin{equation}
\cv= \left(\begin{array}{cc} e^t & 0 \\  0
&e^{-1}\end{array}\right)g(B), \qquad
g(B)= \left(\begin{array}{cc} \1 & 0 \\
-B &\1\end{array}\right)
 \label{vista}
\end{equation}
in terms of the vielbein with $B=0$, and acting on this with the
$O(n,n)$ element $g(B)$ generated by the 2-form $B$. In this way,
the geometry with respect to $\cm$ is the $B$-deformation of $T^n$
and the dual torus $\ti T^n$
 given by
acting with $g(B)$.

\subsection{Equivalence with Standard Formulation}

One approach to solving the constraints is to solve the Bianchi
identities $d\cp =0$ to obtain $2n$ coordinates $X^i, \ti X_i$.
Instead, one can solve $dP=0$ to obtain $n$ coordinates $X^i$
\begin{equation}
P^i_\a=\pa _\a X^i
\label{abc}  \end{equation}
and then solve the self-duality constraint (\ref{cons})
to obtain
$Q_i=\ti Q_i (P, *P, K,*K)$ where
\begin{equation}
\ti Q_i = G_{ij} (*P^j-K^j)+  B_{ij} P^j
\label{abc}  \end{equation}
while (\ref{jcons}) can be solved to give
$J_i= \ti J_i$ where
\begin{equation}
\ti J_i = -  G_{ij} * K^j +  B_{ij} K^j
\label{jtis}  \end{equation}

Then by an argument similar to that given in
\cite{Cremmer:1997ct},  the field equations for both $X$ and $Y$
derived from the doubled lagrangian (\ref{lagd})  together with
the constraints (\ref{cons}),(\ref{jcons}) are completely
equivalent to those derived from the lagrangian
\begin{equation}
\cl _s= \half G_{ij} \,  P^i  \we * P^j + \half B_{ij} \,  P^i  \we   P^j
-G_{ij} \,  P^i  \we K^j
-  \frac{1}{4} G_{ij} \,  K^i  \we * K^j
+ \cl (Y)
\label{lags}  \end{equation}
It is useful to decompose the world-sheet one-form
\begin{equation}
G_{ij} \,  K^j= k_{im}*dY^m- v_{im}dY^m \label{kis}
\end{equation} in terms of some $k_{im}(Y)$, $v_{im}(Y)$.
 Then if $ \cl (Y) $ takes the form
\begin{equation}
\cl (Y)= \half \ti G_{mn} \,  dY^m  \we * dY^n + \half \ti B_{mn} \,  dY^m  \we  dY^n
\label{abc}  \end{equation}
for some $\ti G_{mn} (Y), \ti B_{mn}(Y)$
on the base manifold $N$,
the full lagrangian is
\begin{equation}
\cl (Z)= \half \cg _{PQ} \,  dZ^P  \we * dZ^Q + \half {\cal{B}}_{PQ} \,  dZ^P  \we  dZ^Q
\label{lagsig}  \end{equation}
where $Z^P=(X^i,Y^m)$ are coordinates on the total space of a $T^n$ bundle over $N$
 with
 \begin{eqnarray}
  \cg_{ij}&=&G_{ij}
 \nonumber \\
\cg_{im}&=&k_{im}
 \nonumber \\
\cg_{mn}&=& \ti G_{mn}+ \ha G^{ij} ( k_{im}k_{jn} - v_{im}v_{jn})
\end{eqnarray}
and
 \begin{eqnarray}
   {\cal{B}}_{ij}&=&B_{ij}
 \nonumber \\
 {\cal{B}}_{im}&=&v_{im}
 \nonumber \\
 {\cal{B}}_{mn}&=& \ti B_{mn}+ \ha G^{ij} ( v_{im}k_{jn} - k_{im}v_{jn})
\end{eqnarray}

\subsection{T-Duality Transformation Rules}

Consider an  $O(n,n)$ transformation
by
 \begin{equation}
g=\left(
\begin{array}{cc}
a & b\\
c & d \end{array}\right),
\label{abc}  \end{equation}
where $a,b,c,d$ are $n\times n$ matrices.
This preserves the indefinite metric  $L$, so that
\begin{equation}
g^tLg=L\;\; \Rightarrow \;\;a^t c+c^t a=0,\;\;\;b^t d+d^t
b=0,\;\;\;\ a^t d+c^t b=\1 . \label{abc}  \end{equation} The
transformation rules for $\cm $ (\ref{gtrans})  give  the
non-linear   transformation of $E$ under a T-duality
transformation $g\in O(n,n )$ \cite{rocek},\cite{Giveon:1994fu}
\begin{equation}
E'  = (aE+b)(cE+d)^{-1}. \label{tetrans}  \end{equation} The
transformation (\ref{gtrans}) for $ \cj$ gives, using
(\ref{cjist}), $K'=cJ+dK$. Substituting for $J$ using the solution
$J=\ti J$ (\ref{jtis}) of the constraint (\ref{jcons}) gives
\begin{equation}
{K'}^i= (d^i{}_j + c^{ik} B_{kj}) K^j-c^{ik}G_{kj} *K^j
\label{kttrans}
\end{equation} This then implies the transformation rules for
$v,k$ via (\ref{kis}) and results in  the standard T-duality
transformation rules of
\cite{Buscher:sk},\cite{rocek},\cite{Giveon:1994fu}, provided $\ti
G_{mn},\ti B_{mn}$ are invariant.  This will be seen explicitly in
the next section for $n=1$.

\section{The $n=1$ case}

The simplest example is that of $n=1$, with a fibre that is an $S^1$.
It is instructive to see how things work in this case explicitly.
First, the formalism of the last section is applied to this case, and then
   topology and T-duality  are discussed  for this  example
to motivate the treatment of the general case in sections 4 and 6.

\subsection{The Doubled Formalism for $n=1$}

For the case in which $n=1$, the indices $i,j...$ all take the value $1$,
 so that  $B_{ij}=0$ and $G_{11}=R^2$ where $R(Y)$ is the
 radius of the fibre circle with coordinate $X^1$, with the
identification    $X^1\sim X^1+2\p$. The 2-metric $\cm$ on the
doubled torus is simply
\begin{equation}
\cm= \left(
\begin{array}{cc}
R^2 & 0 \\
0   & R^{-2}
\end{array}
\right)
\label{abc}  \end{equation}
so that the lagrangian (\ref{lagd}) becomes
\begin{equation}
\cl _d= \half
 R^2  dX \we * dX + \half
 R^{-2}  d\ti X \we * d\ti X +
 dX \we *J + d\ti X \we *K+ \cl (Y)
\label{abc}  \end{equation}
and the doubled torus consists of the
$X$ circle of radius $R$ and the dual $\ti X$ circle of radius
$\ti R=1/R$.

Consider the polarization based on picking the   $X$ circle from
$T^2$. The source term is given in terms of $J,K$ with
\begin{equation}
R^2 \,  K = k_{m}*dY^m- v_{m}dY^m
\label{abc}  \end{equation}
and $J=\ti J=-R^{2}*K$.
The equivalent sigma-model is
(\ref{lagsig}) with
 coordinates $Z^P=(X^1,Y^m)$   on the total space of an $S^1$ bundle over $N$
 with
 \begin{eqnarray}
  \cg_{11}&=&R^2
 \nonumber \\
\cg_{1m}&=&k_{m}
 \nonumber \\
\cg_{mn}&=& \ti G_{mn}+ \ha R^{-2} ( k_{m}k_{n} - v_{m}v_{n})
\end{eqnarray}
and
 \begin{eqnarray}
   {\cal{B}}_{11}&=&0
 \nonumber \\
 {\cal{B}}_{1m}&=&v_{m}
 \nonumber \\
 {\cal{B}}_{mn}&=& \ti B_{mn}+ \ha R^{-2} ( v_{m}k_{n} - k_{m}v_{n})
\end{eqnarray}

Consider the $O(1,1;\Z)$ transformation given by the action of
\begin{equation}
g=\left(\begin{array}{cc}
0 & 1 \\
1 & 0
\end{array}\right),
\label{abc}  \end{equation}
In the doubled formalism this takes $\cx \to \cx '=g^{-1}\cx$
so that
\begin{equation}
X'=\ti X, \qquad  \ti X' =X
\label{abc}  \end{equation}
exchanging the two dual circles, while $R'=1/R$ and $J'=K$, $K'=J$.
In the polarized form,
the T-duality
gives $R'=1/R$ and  $K'=\ti J= -R^{2}*K$.
This implies
\begin{equation}
  K' ={R'}^{-2}( k'_{m}*dY^m- v'_{m}dY^m)=-  k_{m}dY^m+v_{m}*dY^m
\label{abc}  \end{equation}
so that
\begin{equation}
k'_{m}= {R}^{-2}v_{m}
\qquad
v'_{m}= {R}^{-2}k_{m}
\label{abc}  \end{equation}
 This, together with the invariance of $\ti G, \ti B$, then gives the standard Buscher T-duality rules
 \cite{Buscher:sk}
 for
 $\cg,  {\cal{B}}$   \begin{eqnarray}
  \cg'_{11}&=& \cg_{11}^{-1}
 \nonumber \\
\cg'_{1m}&=&\cg_{11}^{-1}{\cal{B}}_{1m}
 \nonumber \\
\cg'_{mn}&=& \cg_{mn}
-   \cg_{11}^{-1} ( \cg_{1m}\cg_{1n} -  {\cal{B}}_{1m} {\cal{B}}_{1n})
\end{eqnarray}
and
 \begin{eqnarray}
 {\cal{B}}'_{1m}&=& \cg_{11}^{-1}
\cg_{1m}
 \nonumber \\
 {\cal{B}}'_{mn}&=&  {\cal{B}}_{mn}-  \cg_{11}^{-1} ( {\cal{B}}_{1m}\cg_{1n} -  \cg_{1m}{\cal{B}}_{1n})
\end{eqnarray}

\subsection{T-Duality and Transition Functions for $n=1$.}

Over a point $Y\in N$, there is a $T^2$ consisting of the $X$ circle with
radius $R(Y)$ and the $\ti X$
 circle with radius $\ti R(Y)=1/R$, and there is a choice of polarization,
 i.e. a choice of which one-cycle is part of the space-time and which is the dual circle.
Suppose we choose  the $X$  circle  to be the one that is part of the
physical spacetime. T-duality acts by taking $R \to 1/R$, so that
the physical $X$ circle now has radius $1/R$ and the dual $\ti X$
one has radius $R$. However, there is an equivalent way of the
viewing the T-duality transformation: one could keep the radius
$R$ fixed, but change the polarization so that it is now the $\ti
X$ circle of radius $1/R$ that is part of the physical spacetime.
The T-duality can then be viewed as an active transformation
transforming the geometry of the $T^2$ with $R \to 1/R$, or as a
passive transformation in which the $T^2$ geometry is kept fixed,
but the choice of polarization is changed. In either case, the
physical spacetime is changed from one with a circle of radius $R$
to one with a circle of radius $1/R$. As the   conformal field
theory   for a circle of radius $R$ is the same as that for a
circle of radius $1/R$, the change does not affect the physics,
but changes the variables used to describe it.

Consider first the case in which  the spacetime is a
 trivial bundle $S^1\times N$, so that the doubled space is
 $T^2\times N$.
The physical spacetime is   $S^1\times N$ embedded in $T^2\times
N$ and T-duality acts to \lq rotate' the $S^1$ within
$T^2=S^1_R\times S^1_{1/R}$ from the $S^1_R$ of radius $R$ to the
$S^1_{1/R}$ of radius $1/R$. In the active view, the direction of
the polarization is kept fixed but the $T^2$ is rotated, while in
the passive view the $T^2$ is kept fixed but the polarization is
rotated -- the polarization is initially a vector pointing in the
$X$ direction in the $X,\ti X$ plane and the same effect is
obtained by either rotating the vector through $\p/2$  while
keeping the $X,\ti X$ axes fixed, or by keeping the vector fixed
and rotating the axes through $-\p /2$.

Next   consider a situation with a non-trivial transition
function. Suppose $U,U'$ are two coordinate patches in $N$ with
non-trivial intersection and that the corresponding  patches of
$M$ are $U\times S^1_R$ with a circle of radius $R(Y)$ and
$U'\times S^1_{R'}$ with a circle of radius $R'(Y)$. These can be
glued together to form (part of) a manifold if $R(Y)=R'(Y)$ for
$Y\in U\cap U'$. However, they can be glued together with a
T-duality transition function to form (part of) a T-fold if
$R'(Y)=1/R(Y)$ for $Y\in U\cap U'$. The T-fold transition clearly
does not make a smooth space as it involves gluing a large circle
to a small one, but from the conformal field theory point of view
it is the same CFT over $U$ and over $U'$ and the transition is
simply a change of the variables used to parameterise the CFT,
changing which of $\cp ^1,\cp ^2$ is to be viewed as a momentum
and which is to be viewed as a winding number.

In the doubled formalism the two patches become $U\times
S^1_R\times S^1_{1/R}$ and $U'\times S^1_{R'}\times S^1_{1/R'}$.
In the manifold case with $R=R'$, these can be glued together in
the natural way, gluing the $X$ circle to the $X'$ one, and the
$\ti X$ circle to the $\ti X'$ one. However, for the T-fold case
with $R'=1/R$, there is now the possibility of making a manifold
by gluing the $X$ circle to the $\ti X'$ one as both have radius
$R$, and gluing the  $\ti X$ circle to the $  X'$ one as both have
radius $1/R$. This fits together to make a smooth manifold $(U\cup
U')\times T^2$, where over each point in $U\cup U'$ there is a
circle of radius $R$ and one of radius $1/R$. However, the
physical spacetime has two patches, $U\times S^1_R $ over $U$ with
the $X$ circle of radius $R$ and  $U'\times S^1_{R'}$ over $U'$
with the $X'$ circle of radius $R'=1/R$, and clearly these patches
do not fit together to form a submanifold of the doubled space
$(U\cup U')\times T^2$. Thus the polarization changes by a
T-duality in going from  $U$ to $U'$. There are local patches of
spacetime in which there is a conventional picture of the physics
in terms of strings moving in that spacetime, but there  is no
global spacetime. There is a global doubled space $(U\cup
U')\times T^2$, but the physical  subspace is defined locally and
jumps discontinuously through T-duality transition functions.

This generalises to the case $n>1$. In the next section, it will
be seen that in general T-duality acts to change the polarization
by changing the physical subspace $T^n \subset T^{2n}$, while in
section 6 it will be seen that over each patch $U\subset N$ there
is a local patch of spacetime $U\times T^n$ embedded in the
doubled space $U\times T^{2n}$, but for T-folds these do not fit
together to form a spacetime manifold, even though there is a
doubled manifold which is a $T^{2n}$ bundle over $N$.

\section{T-Duality Symmetry  }

The doubled lagrangian $\cl _d$ is invariant under the $O(n,n)$ transformations (\ref{gtrans}),(\ref{xtrans}):
\begin{equation}
\cl_d(\cm',\cj', \cx')=\cl_d(\cm,\cj, \cx) \label{abc}
\end{equation} This means that $\cl_d(\cm,\cj, \cx)$ and
$\cl_d(\cm',\cj', \cx)$ define the same theory, as they are
related by a field redefinition $\cx \to g\cx $. Part of the
specification of the   theory is   a choice of $(\cm,\cj)$, and
choices $(\cm,\cj)$, $(\cm',\cj')$ related by $O(n,n;\Z)$
transformations give  equivalent doubled theories.

A further piece of data needed to specify the theory is a choice
of polarization, i.e. a choice of a projector $\P ^i{}_I$ which
selects a null submanifold  $ T^n$ of the doubled space $T^{2n} $
as the physical spacetime, together with the complementary projector
$\ti \P_{iI}$ onto $\ti T^n$. Two projectors $\P, \P'$ define the
same $T^n$ if they are related by a diffeomorphism of $T^n$, so
that
\begin{equation}\label{zda}
    \P' =h \P
\end{equation}
for some $h^i{}_j \in GL(n,\Z)$.
 Then $(\cm,\cj,\P)$,
$(\cm',\cj', \P')$ define equivalent theories provided that the
projector is transformed as
\begin{equation}
\P'= h \P g \label{abc}  \end{equation} for some $h\in GL(n,\Z)$.
That is, if one simultaneously transforms the geometric data
$(\cm,\cj)$ and the polarization $\P$, then nothing changes. As we
shall now show, the conventional form of a T-duality
transformation consists of keeping $\P$ fixed and transforming
$(\cm,\cj)$
 according to the Buscher rules. However, the same effect can be obtained by keeping
 $(\cm,\cj)$ fixed and transforming the polarization $\P$.
 Thus T-duality can be thought of in two ways. In the first, the doubled torus is transformed, but the
 projection onto the physical subspace is kept fixed, while in the second the doubled torus
 is kept fixed, but the choice of physical subspace is changed.

 It will be useful to introduce the notation
 $\hat I$ for the $O(n,n)$ indices in the $GL(n)$ basis, so that
for  any vector $v$, $v^{\hat I}=(v^i,v_i)$ and the matrix giving
the change from an arbitrary basis to the $GL(n)$ basis is
\begin{equation}
\Phi^{\hat I} {}_J= \begin{pmatrix} \P ^i{}_J \\  \ti \P  _{iJ}  \end{pmatrix}
\label{abc}  \end{equation}
 with the corresponding matrix for  the
dual representation
$\hat \Phi = L^{-1} \Phi L$ so that
\begin{equation}
\hat \Phi _{\hat I} {}^J= \begin{pmatrix} \ti \P  _{i}{}^J \\
\P ^{iJ }\end{pmatrix} \label{abc}  \end{equation} where $  \P
^{iJ }= \P ^i{}_I L^{IJ}$, $\ti \P  _{i}{}^J=\ti \P  _{iI}L^{IJ}$.
The matrix $\Phi^{\hat I} {}_J$ can be thought of as a
representative of the coset $O(n,n)/GL(n,\R)$, or as a \lq
vielbein' converting $O(n,n)$ indices to $GL(n)$ ones. (The
context should avoid confusion between the vielbein $\Phi$ and the
dilaton.)
It will also be useful to introduce
\begin{equation}
\car^{\hat I} {}_J= \begin{pmatrix} \P ^i{}_J \\  -\ti \P  _{iJ}  \end{pmatrix}
\label{abc}  \end{equation}
so that in the $GL(n)$ basis $\car^2=\1$; this will play a useful role in the following sections.

Then the equations giving components in the $GL(n)$   basis can be
written as
\begin{equation}
\Phi  \cp =
\left( \begin{array}{c} P  ^i  \\ Q _i  \end{array}\right), \qquad
\Phi \cj   =\left( \begin{array}{c}J_i      \\ K^i  \end{array}\right)
\label{jfiss}
  \end{equation}
and
\begin{equation}
\hat \Phi \cm \hat \Phi ^t=  \left(
\begin{array}{cc}
G-BG^{-1}B & BG^{-1} \\
-G^{-1}B   & G^{-1}
\end{array}
\right).
\label{hfis}
  \end{equation}
and this notation will help in    following  the effects of changes of polarization explicitly.
In particular, (\ref{hfis}) defines a metric $G_{ij}$ and 2-form $B_{ij}$ in terms of $\cm$ and
a polarization $\Phi$
 \begin{eqnarray}
  G^{ij}&=& \P ^{iI}\P^{jJ} \cm _{IJ}
  \nonumber \\
 B_{ij}G^{jk}&=& \ti \P _i{}^{I}\P^{kJ} \cm _{IJ}
 \end{eqnarray}
Similarly, $K_i$ is defined from (\ref{jfiss}) by
\begin{equation}
K^i= \P^{i J} \cj _J
\label{abc}  \end{equation}

The T-duality transformation rules $G\to G'$, $B\to B'$, $K\to K'$
(\ref{tetrans}), (\ref{kttrans}) are then obtained   using the
$O(n,n)$ transformations for $\cm, \cj$ while keeping the
polarization $\Phi$ fixed,
\begin{equation}
\cm \to \cm '= g^t \cm g,  \qquad \cj \to \cj ' = g^t \cj,  \qquad
\Phi \to \Phi'= \Phi \label{hact}  \end{equation}
 so that
\begin{eqnarray}
  G^{-1}&=& \P \cm  \P^{t} \quad  \to \quad  (G')^{-1}=\P  g^t \cm g  \P^{t}
  \nonumber \\
 B G^{-1}&=& \ti \P \cm  \P^{t}  \quad \to \quad B'(G')^{-1}= \ti \P  g^t \cm g  \P^{t}
 \nonumber \\
 K&=& \P \cj   \quad \to \quad K' =  \P  g^t \cj
 \end{eqnarray}
These same transformations $G\to G'$, $B\to B'$, $K\to K'$ can also be obtained by keeping $\cm$ fixed while transforming $\Phi$
\begin{equation}
\cm \to \cm '=  \cm ,  \qquad \cj \to \cj ' =   \cj,  \qquad \Phi
\to \Phi'= \Phi g \label{hpas}  \end{equation} so that
\begin{equation}
 \P  \to \P '= \P  g   \qquad  \ti \P  \to\ti  \P '= \ti  \P  g
\label{ppas}  \end{equation}
Note that this could be supplemented with a $GL(n)$ transformation, so that
$\P  \to \P '= h^{-1}\P  g  $,  $ \ti \P  \to\ti  \P '=(h^t)^{-1} \ti  \P  g $.

Thus the T-duality transformations can be viewed either as active
transformations in which the geometry $\cm, \cj$ is changed while
$\P$ is kept fixed (\ref{hact}), or as a passive one in which  the
geometry $\cm, \cj$ is kept fixed but the polarization is changed
(\ref{hpas}),(\ref{ppas}). In the latter viewpoint, the doubled
geometry is unchanged, but the choice of physical subspace is
transformed. The symmetry under T-duality is then the statement
that the physics does not depend on the choice of physical
subspace.

\section{Product Structures and Polarizations}

The polarization can be characterised in terms of a product
structure on the  $T^{2n}$ fibres. A local product structure (sometimes called a
real structure or pseudo-complex structure; see e.g. \cite{Gates:nk}) is a tensor $\car
^I{}_J$ satisfying
\begin{equation}
\car ^2 =\1 \label{prods}  \end{equation} and also satisfying an
integrability condition analogous to the vanishing of the
Nijenhuis tensor for a complex structure.
The local product structure allows the construction of two projectors, $\ha (\1 \pm
\car)$. A product structure defines a splitting of the tangent
space into two eigenspaces of $\car$ with eigenvalue $\pm 1$. For a torus
$T^{2n}$, this will extend to a splitting of the periodic torus coordinates  into those of two $T^n$
eigenspaces if the product structure is {\it{integral}}
\begin{equation}\label{asd}
    \car \in GL(2n,\Z)
\end{equation}
so that it acts on the coordinates while preserving their
periodicities. 

A metric $L_{IJ}$ is {\it{pseudo-hermitian}} with respect to $\car$ if
\begin{equation}
L_{IK}\car ^K{}_J + L_{JK}\car ^K{}_I=0 \label{pherm}
\end{equation}A choice of polarization of $T^{2n}$ then
corresponds to a choice of real structure with respect to which the $O(n,n)$ metric 
$L_{IJ}$ is
pseudo-hermitian  and which is integral, i.e. it satisfies (\ref{asd}).
Then  the projector $\P$ onto the physical  space is  $\ha (\1 + \car)$ and
 the projector
$\ti \P$ onto the dual  space is  $\ha (\1 - \car)$.
A product
structure and pseudo-hermitian
$O(n,n) $ invariant metric
are together preserved by the subgroup $GL(n,\R) \subset
O(n,n)$, and for the transformations acting on the torus, it is preserved by
$GL(n,\Z) \subset
O(n,n;Z)$.

Under an active T-duality transformation $\cm,\cj$ change as in  (\ref{hact}) but $\car$ does not, while under a
 passive T-duality transformation
 $\cm,\cj$ do not change but $\car $ transforms as
 \begin{equation}
  \car
\to \car'= g^{-1}\car g  \label{rhpas}  \end{equation}

A second local product structure is defined by
\begin{equation}
\cs ^I{}_J = L^{IK}\cm _{KJ} \label{abc}  \end{equation} which
satisfies $\cs ^2 = \1$, which follows from $\cm=\cv ^t \cv$ with $\cv\in O(n,n)$.
This satisfies
\begin{equation}
L_{IK}\cs ^K{}_J - L_{JK}\cs ^K{}_I=0 \label{abc}
\end{equation}
The sign here implies the metric is not pseudo-hermitian with
respect to $\cs$, but that the metric is {\it compatible} with the
product structure, so that locally the metric is a product metric. Note that also
$\cm$ is compatible with $\cs$, 
\begin{equation}
\cm_{IK}\cs ^K{}_J - \cm_{JK}\cs ^K{}_I=0 \label{abc}
\end{equation}
and the product
\begin{equation}
\ci = \cs\car
\label{abc}  \end{equation}
is a complex structure
\begin{equation}
\ci ^2 =-\1
\label{abc}  \end{equation}
with respect to which the metric  $L$ is hermitian
\begin{equation}
L_{IK}\ci ^K{}_J + L_{JK}\ci ^K{}_I=0 \label{abc}
\end{equation}
The  three matrices
$q_a$ with $q_1=\car, q_2=\cs,q_3=\ci$
satisfy the
pseudo-quaternionic algebra
\begin{equation}
q_a q_b = f_{ab}{}^c q_c + \eta _{ab} \label{abc}  \end{equation}
where $a,b=1,2,3$,   $ f_{ab}{}^c $ are the structure constants of
$SL(2,\R)$ and $ \eta _{ab}= diag (1,1,-1)$ is the Cartan metric
of $SL(2,\R)$. The  commutation relations are those of the Lie
algebra $SL(2,\R)$ and the subgroup of $O(n,n)$ preserving these
three structures is $O(n)$ (the diagonal subgroup of   $O(n)\times
O(n)$).

A polarization requires an integral product structure $\car$
satisfying (\ref{asd}), which requires that $\car$ is constant,
while $\cs$ depends on the moduli $\cm(Y)$ or $G(Y),B(Y)$,  so
that $\cs,\ci$ are tensors which   depend on the coordinates $Y$,
and $\cs$ is not integral.

Then there are tensors $\car,   L$ which are
  constant matrices in the adapted  coordinate system used here, and in a suitable basis they  take the form
\begin{equation}
\car^I{}_{J}=\left(\begin{array}{cc}
\1 & 0 \\
0&-\1
\end{array}\right),
\qquad
L_{IJ}=\left(\begin{array}{cc}
0 & \1 \\
\1 & 0
\end{array}\right).
\label{sdfdsf}
 \end{equation}
 In addition there are $\cm_{IJ},\cs^I{}_J, \ci^I{}_J$ which depend on $Y$.
 Choosing coordinates so that, at some point $Y_0$, $\cm_{IJ}(Y_0)=\d_{IJ}$,
 then at the point $Y_0$ in a suitable basis the  $\cm_{IJ},\cs^I{}_J, \ci^I{}_J$
 become
 \begin{equation}
\cs^I{}_{J}=\left(\begin{array}{cc}
0 & \1 \\
\1&0
\end{array}\right),
\qquad
\ci^I{}_{J}=\left(\begin{array}{cc}
0 &- \1 \\
\1&0
\end{array}\right),
\qquad
\cm_{IJ}=\left(\begin{array}{cc}
\1 & 0 \\
0 & \1
\end{array}\right),
\label{sdfdsf}
 \end{equation}
and it is straightforward to check that these matrices satisfy the algebraic conditions given above.

The projection $\P:T^{2n} \to T^n$
leads to a fibration of $\ti M$ over $M$ in the case in which
 $M$ is a geometric background, with $\P:(X,\ti X, Y) \to (X, Y)$.
 It also leads to a projection from the tangent bundle of
 $T^{2n}$ to the sum of the tangent and cotangent bundles of
 $T^n$, $\P: T(T^{2n}) \to (T\oplus T^*)(T^{n})$ with
 $\P:(X^i,\ti X_i, P^i, \ti P_i) \to (X^i, P^i, \ti P_i)$.
 In the tangent space $\R^{2n}$
at a point of $T^{2n}$, the choice of a projector $\P$ is the choice of a
subgroup $GL(n,\R)\subset O(n,n)$, which is the choice of a maximal null
subspace or maximally isotropic subspace, which corresponds to choosing a
pure spinor of $Spin(n,n)$ for each point on $T^{2n}$ \cite{Hitchin}.
Then projecting onto $M$, there is a pure spinor of $Spin(n,n)$
for each point in $M$,
 and so there is a real analogue of  the generalised Calabi-Yau structure of  \cite{Hitchin}
  on the $T^n$ fibres of $M$ (preserved by $GL(n)\subset O(n,n)$ instead of by $U(m,m)\subset O(2m,2m)$).
Each of the structures $q_a$ defines a generalised complex or real structure for
the fibres $T^{n}$ of $M$ and together furnish what might be called a generalised pseudo-hyperkahler structure.

\section{Topology}

Let $\{U_\a\} $ be an open cover of the base $N$, $N= \cup _\a
U_\a$\footnote{In this section $\a,\b$ will label coordinate patches and not
world-sheet coordinates.}, and suppose that there is a set of  transition functions
$\ti g_{\a\b}$ satisfying the usual compatibility relations
\begin{eqnarray}
\ti   g_{\a\a}&= \1 \qquad &  in \quad U_\a
  \nonumber \\
\ti g_{\a\b}&= \ti g_{\b\a} ^{-1}  \qquad &  in \quad U_\a \cap U_\b
 \nonumber \\
\ti g_{\a\b} \ti g_{ \b\g} \ti g_{\g\a } &= \1  \qquad   &  in \quad U_\a \cap U_\b\cap U_\g
\label{compat}
 \end{eqnarray}
If the transition functions are in $GL(2n;\Z)\ltimes U(1)^{2n}$ they can be used to
construct a space $\ti M$ as a $T^{2n}$ bundle over $N$. The
coordinates $\cx ^I$
    on $T^{2n}$  take values in $\R^{2n}/ \Gamma$ where $\Gamma$ is a  lattice
   with automorphism group  $GL(2n;\Z)$.
Then the coordinates in $U_\a\times T^{2n}$ are $(Y_\a, \cx^I_\a)$
and in $U_\a \cap U_\b$ the $T^{2n}$ fibre coordinates are related
by a matrix $(g_{\a\b})^I{}_J$ in $GL(2n;\Z)$ and a shift $x_{\a\b}$ in $U(1)^{2n}$
 \begin{equation}
\cx_\a = g_{\a\b}^{-1}\cx_\b +x_{\a\b} \label{xpa}  \end{equation} and  the
transformations  for $\cp, \cm, \cj,
L$ in overlapping patches  given by
\begin{equation}
\cp_\a = g_{\a\b}^{-1}\cp_\b , \qquad L_\a =
g_{\a\b}^{-1}L_\b (g_{\a\b}^{-1})^t
\label{jpa}  \end{equation}
and
\begin{equation}
\cm_\a = g_{\a\b}^{t}\cm_\b g_{\a\b} , \qquad
\cj_\a = g_{\a\b}^{t}\cj_\b
\label{mpa}  \end{equation}
are those following from the T-duality rules. The transition
functions can be thought of as defining a change of coordinates on
the doubled torus in transforming from $U_\a\times T^{2n}$ to
$U_\b\times T^{2n}$, and these transformations of $\cp, \cm, \cj,
L$ are those needed for them to be well-defined objects on $\ti
M$; such objects will be referred to as \emph{tensors} on the
doubled bundle.

For the special class of bundles for which $g_{\a\b}\in O(n,n;\Z)
\subset GL(2n;\Z)$, $L_\a=L_\b$ and there is a constant
 metric $L_{IJ}$ of signature $(n,n)$. The positive
definite metric $\cm$ depends on $Y$ in general and will transform
non-trivially between patches. If $N$ is orientable, the total
space $\ti M$ will be orientable only if the structure group is in
$SL(2n;\Z)$, which for invariant $L$ requires the structure group
to be in $SO(n,n;\Z)$. For structure group
$O(n,n;\Z)$, $\ti M$ will be non-orientable in general.

Next we wish to discuss a  polarization defined by   a product structure $\car$
satisfying (\ref{prods}),(\ref{pherm}),(\ref{asd}).
Suppose such a structure $\ti \car_\a $  satisfying (\ref{prods}),(\ref{pherm}),(\ref{asd}) is
introduced for  each patch $U_\a$.
As it is integral,  $\ti \car_\a $ must be   constant over $U_\a$, but
 might be different in different patches ($\ti \car_\a  \ne \ti \car_\b $ for $\a \ne \b$).
Then  $\ti \car_\a $ defines a splitting of the fibres over
$U_\a$, $T^{2n} \to T^n \oplus \ti T^{n}$, selecting a $T^n$ and a dual $\ti T^n$.
The $\ti \car_\a$ will form a tensor if
they are related by the geometric transition functions
 \begin{equation}
\ti \car_\a =g_{\a\b}^{-1}\ti \car_\b g_{\a\b}
\label{rpa}
\end{equation}
If this is the case, the $T^n$ subspaces of the fibres selected  by restricting to the
$+1$ eigenspace of $\car$
fit together to form a $T^n$ bundle over $N$, as do the dual
 tori $\ti T^n$. Given a choice of $T^n$ subspace over a patch $U_{\a_0}$ corresponding to a
product structure  $\ti \car _{\a_0}$, then
defining $\ti \car $ in all other patches through (\ref{rpa}) gives a
 choice of $T^n$ for all other patches and  gives a $T^n$ bundle over $N$.
The subgroup of $O(n,n;\Z)$ preserving a  product structure (with a
pseudo-hermitian $O(n,n) $ metric) is $GL(n,\Z)$, so that
if the structure group of the bundle is  $GL(n,\Z)$, then $\ti  \car_\a = \ti \car_\b$.

The theory is specified by a choice of $(\cm,\cj,\car)$ for each patch,
 and if all three are tensors with geometric transition functions
 (\ref{mpa}),(\ref{rpa}), then the polarization selects a physical
 subspace $U_\a \times T^n$ for each $\a$ and these fit together to form a $T^n$ bundle over $N$.
However, such geometric transition functions are not T-duality
transition functions. Recall that a T-duality corresponds to
transforming $(\cm,\cj)$ while keeping $\car$ fixed, or  equivalently to
transforming $\car$  while keeping $(\cm,\cj)$ fixed. The
construction here involves a bundle with $(\cm,\cj)$ glued
together with $O(n,n;\Z)$ transition functions  (\ref{mpa}), so
that for the transition functions to be   T-dualities requires a
constant fibre product structure with trivial transition functions
\begin{equation}
 \car_\a = \car_\b \label{triv}
\end{equation}
Then the T-fold is a bundle with  $O(n,n;\Z)$
transition functions $g_{\a\b}$   with $(\cm,\cj,\car)$ glued  using
 (\ref{mpa}),(\ref{triv}), and for each $\a$, this selects a physical
 spacetime patch $U_\a \times T^n$.

A constant product structure (\ref{triv})  will be consistent
with the geometric transition functions  (\ref{rpa}) only if the
transition funcitons $g_{\a\b}$ are all in the geometric subgroup
$GL(n,\Z)\subset O(n,n;\Z)$ preserving $\car$. Then for \lq
geometric' bundles with $GL(n,\Z)$ structure group, the physical
subspace selected by a polarization corresponding to a constant
product structure (\ref{triv}) is a $T^n$ bundle over $N$, and the
local   patches $U_\a \times T^n$ fit together to form a spacetime
manifold, while for $O(n,n;\Z)$ structure group they do not.

Any spacetime point is in a local spacetime
patch $U_\a\times T^{n}$ and the local physics is described
by strings moving in this spacetime patch.
For a geometrical
background, these   fit together to form a manifold which is a
$T^n$ bundle over $N$, but for a non-geometric background or a
T-fold these local patches do not fit together to form a manifold.
For a geometric background there is a global polarization, i.e.
there is a well-defined way of choosing a physical spacetime
inside the doubled space, while for a non-geometric one a physical
spacetime can only be found locally, but a global one does not
exist.

This can also be thought of from a \lq passive' viewpoint. Given a
choice of polarization corresponding to a product structure
$\car$,    choose a patch $U_{\a_0}$, say, and introduce a product
structure    $\ti \car_{\a_0}=\car$ in this patch and extend $\ti
\car$ to all patches using the   rule  (\ref{rpa}) to define a
tensor  $\ti \car_{\a } $, and there will be $g_{\a_0\b}\in
O(n,n;\Z)$ such that
\begin{equation}
\ti \car_{\b }
=g_{\b\a_0}^{-1}\ti \car_{\a_0} g_{\b\a_0}
\end{equation}
Then $\ti \car_{\a } $ defines a $T^n$  bundle over $N$,
and selects a reference $T^n$ over each patch $U_\a$.
Then the  physical subspace is defined by $\car$ and is obtained by
an $O(n,n;\Z)$ transformation of the reference $T^n$ subspace, as in $U_\b$
\begin{equation}
  \car
=g_{\b\a_0}\ti \car_{\b} g_{\b\a_0}^{-1}
\end{equation}

Over each $U_\a$, there is a physical space $U_\a\times T^n$
embedded in $U_\a\times T^{2n}$ which can be thought of as a piece
of the physical spacetime. If the transition functions are
geometric (in the appropriate $GL(n;\Z)$ subgroup), these pieces
fit together to form a spacetime which is a $T^n$ bundle over $N$,
but  more generally the local pieces of spacetime do not fit
together to form a manifold.

T-duality acts on these bundles by an $O(n,n;\Z)$ transformation
$g$ which is constant, so that it is the same for each patch $\a$.
It   takes the physical subspace defined by a product structure  $\car$ to
the physical subspace corresponding to $\car'=g^{-1} \car g$ while taking
the group $GL(n,\Z)$ of elements $U$ preserving $\car$ to a conjugate $GL(n,\Z)$ subgroup
of $O(n,n;\Z)$ consisting of elements $gUg^{-1}$. In this way it can change the maximal null
$T^n\subset T^{2n}$ to any other maximal null $T^n$ subspace and,
as T-duality is a symmetry, different choices of polarization  give the same
physics, but in T-dual representations.

\section{Open Strings and D-Branes}

The doubled formalism can be used to discuss open strings in $M$ also. Consider first
the   case of a trivial bundle with $M=N\times T^n$.  If, for
some $i$, $X^i(\s^\a)$ is an open string coordinate  satisfying Neumann boundary conditions $\pa _n X^i=0$ at
a world-sheet boundary, then the T-dual coordinate $\ti X_i$ will
satisfy Dirichlet boundary conditions $\pa _t \ti X_i=0$, where
$\pa _n$ and $\pa _t$ are the world-sheet derivatives normal and
tangential to the boundary, respectively.
Conversely, if $X^i$ is Dirichlet, $\ti X_i$ will be Neumann.
 Then the $2n $ doubled
coordinates $\cx^I$ split into $n$ Neumann directions $\cx _N $ and
$n$ Dirichlet directions $\cx_D$. The Dirichlet directions form a
maximal null $T^n$ submanifold.

In addition, a polarization gives  a splitting of the $\cx^I$ into
$n$ physical coordinates $X$ and $n$ dual coordinates $\ti X$, and
in general this will be different from the splitting according to
boundary conditions. Then the $n$ physical coordinates $X$ split
into   $p_X$ Dirichlet coordinates $X_D$ and $n-p_X$ Neumann
ones $X_N$ for some $p_X$, while the dual coordinates $\ti X$
split into  $n-p_X$ Dirichlet coordinates $\ti X_D$ and $p_X$
Neumann ones $\ti X_N$, so that the $n$ Dirichlet coordinates are
$\cx_D=(X_D,\ti X_D)$ and
 the $n$ Neumann coordinates are
$\cx_N=(X_N,\ti X_N)$.
In the physical $T^n$, there is a D-brane wrapping the $T^{p_X}$ with coordinates $X_D$.
If $p_Y$ of the
coordinates of $N$ satisfy Dirichlet boundary conditions, then
there is a D$p$-brane with $p=p_X+p_Y$ wrapping a $T^{p_X}$  submanifold of
  the internal $T^n$.

  In the doubled formalism, there is a D$p$-brane  with $p=p_Y+n$  wrapping the $T^n$
submanifold (with coordinates $\cx_D$) of
  the internal $T^{2n}$,
   and how much of this D-brane is visible in the
physical spacetime depends on how many of the $p_Y+n$ Dirichlet
directions are in the physical slice.
 A T-duality will change the physical polarization to a different
 $T^n\subset T^{2n}$ so that the dimension of the intersection with the Dirichlet $T^n$ will change,
 and   the number $p_X$ of physical Dirichlet directions will also
 change.
For example, for a spacetime with a $T^2$ fibration, $n=2$, the
doubled fibre coordinates consist of two Dirichlet directions
$\cx_D^1,\cx_D^2$ and two Neumann ones $\cx_N^1,\cx_N^2$. If the physical
polarization consists of the two Dirichlet directions
$X=(\cx_D^1,\cx_D^2)$, this represents a $D(p_Y+2)$ brane wrapped on the internal
$T^2$, while a T-duality in the $\cx_D^2$ direction will give
a physical spacetime with coordinates $\cx_D^1,\cx_N^2,Y$ with a
$D(p_Y+1)$ brane wrapping the $\cx_D^1$ direction. A further
T-duality gives a physical spacetime with coordinates
$\cx_N^1,\cx_N^2,Y$ and there is a $Dp_Y$ brane that doesn't wrap the
internal torus.

Consider now the interacting case, with lagrangian (\ref{lagd}).
The doubled torus coordinates $\cx ^I$ are to be split into coordinates
  satisfying Neumann boundary conditions on the edge of the world-sheet, and coordinates
 satisfying Dirichlet ones.
 Introducing a Dirichlet projector
 $(\P_D)^I{}_J$
 and a Neumann projector  $(\P_N)_I{}^J$
 the   boundary conditions are
 \begin{equation}
\P_D \pa _t\cx=0
\end{equation}
and
\begin{equation}
\P_N   ( \cm \pa_n  \cx+ \cj _n)=0
\end{equation}
on the boundary.
Varying the action with lagrangian (\ref{lagd}) gives a boundary term that will vanish if
\begin{equation}
\P_D \P_N^t=0
\end{equation}
The boundary conditions are consistent with the constraint (\ref{cons}) only if 
\begin{equation}
\P_D  L =L \P_N
\end{equation}
(with $L=L^{IJ}$ the inverse metric).
It then follows that there is a suitable choice of coordinates such that
\begin{equation}
L^{IJ}=\left(\begin{array}{cc}
0 & \1 \\
\1& 0
\end{array}\right), \qquad (\P_D)^{I}{}_J=\left(\begin{array}{cc}
\1& 0 \\
0& 0
\end{array}\right),
\qquad (\P_N)_I{}^{J}=\left(\begin{array}{cc}
0 & 0 \\
0 & \1
\end{array}\right)
\label{dfssfsdh}
 \end{equation}
The boundary conditions in $N$ will be the standard ones,
corresponding to a D$p_Y$-bane embedded in $N$, for some $p_Y$.

The Dirichlet coordinates $\cx_D=\P_D \cx$ are then coordinates for a maximal null $T^n$ within $T^{2n}$.
For each patch $U_\a \subset N$, there will be  Dirichlet and Neumann projectors
$(\P_D)_\a,(\P_N)_\a$ and so there will be a corresponding local product structure
$(\hat \car _\a )^I{}_J$.
Consistency of the boundary conditions requires that
the $\hat \car _\a$
  satisfy a condition of the form
(\ref{rpa}) so that the Dirichlet fibres fit together to make a
$T^n$ bundle over $N$. In the doubled picture, one could say that
there is a D-brane wrapping the $T^n$ fibres of this Dirichlet
bundle. Over each patch  $U_\a$ there is a $T^{2n}$, and two $T^n$
subspaces,  the Dirichlet $T^n$ (parameterised by $\cx_D$, with
complement the Neumann $T^n$) and the physical $T^n$
(parameterised by $X$, with complement the dual $\ti T^n$), and
those physical directions $X_D$ which are also Dirichlet will be
where the physical D-brane is said to be. The intersection of the
Dirichlet $T^n$ and physical  $T^n$ will be a $T^{p_X}$
(parameterised by $X_D$) and there is a physical D-brane wrapping
the $T^{p_X}$ subspace of the physical $T^n$, so that there is a
$D(p_X+p_Y)$-brane in the patch of the physical space $U_\a\times
T^n$. A T-duality will change the physical $T^n$ subspace and so
change $p_X$. The transition function between two intersecting
patches on $N$ will be a T-duality changing the physical space and
hence the value of $p_X$ in general, unless the transition
functions are in the geometric $GL(n;\Z)$. Thus for a T-fold, a
D$p$-brane in one region can become a D$p'$-brane in another with
$p'\ne p$. For example, for a twisted reduction with T-duality
monodromy, taking a D$p$-brane round the reduction circle will
bring it back transformed by the T-duality.

\section{Discussion}

For the theories discussed here, at each point in $N$  there is an internal CFT  with $2n$ currents $\cp ^I$ and associated conserved charges. The CFT can be represented by a sigma-model with a torus $T^n$ as target space, in which $n$ of the charges are interpreted  as momenta on the $T^n$ and the remaining $n$ are   interpreted as string winding charges.
However, this is not unique as the same CFT can be represented in terms of different tori, depending on how the $\cp$-charges  are split into momenta and string charges. Locally, one can pick a polarization of the $\cp$, and so choose which of the $n$-torus subspaces is to be viewed as the internal part of the spacetime. Often there will be a natural choice of torus in which all the radii are large compared to the string length so that the low-energy field theory description is useful.
However, for T-folds, there is no global choice   of polarization and so over different regions in the base space $N$ there will be different choices of torus representation of the CFT.
In such cases, there is typically no representation that has a
low-energy  description as a field theory on the total space, but on dimensionally reducing on the internal space, one obtains a   low-energy  description as a field theory on the base $N$.

For each region on $N$, some of the  $\cp ^I$   are geometrical and realised as momenta, but 
the subset of the  $\cp ^I$ that is
  geometrical  can change from region to region,
and each of the $\cp^I$ is potentially  geometrical.
In the doubled formalism, all are treated as geometrical and correspond  to  momenta on the doubled torus $T^{2n}$ so that the duality is a manifest symmetry, and the conventional description arises from choosing a  $T^n$ subspace that is to be interpreted as part of the physical spacetime.
The duality twists in the transition functions mean that there is no way to separate the spacetime momenta from
string winding states, and so both must be treated on the same footing, and the doubled formalism provides a natural way of doing so.

The discussion  has been largely classical and it is important to
extend this  to the quantum theory. First the theory should be
coupled to a curved world-sheet and the Fradkin-Tseytlin  coupling
of the world-sheet curvature to    the dilaton $\Phi(Y)$ is added
(the dilaton does not depend on $\cx$ as T-duality requires
isometries in the torus directions). The conventional approach to
quantization would be to choose a polarization of $\cx$ into
$X,\ti X$, solve the constraint (\ref{cons}) to express $\ti X$ in
terms of $X$ and so obtain a sigma-model formulation which can be
quantized in the conventional way, with consistency requiring
conformal invariance, and the change in the functional measure
under T-duality gives the dilaton transformation (\ref{fitrans}).
However, if there is not a global way of choosing a polarization,
then there is an obstruction to doing this globally. One approach
would be to use this polarized form locally and try and take into
account the T-duality transition functions, but this would make
many features of the theory obscure.

It seems natural to  use instead the duality-covariant doubled
formulation,  integrating over the $\cx$. In this approach the
difficulty lies in implementing the self-duality constraint
(\ref{cons})  in the quantum theory. The constraint (\ref{cons})
says, roughly speaking,  that half of the $\cx$ are right-movers
and half are left-movers. There are several  approaches that have
been proposed for the quantization of  chiral bosons
% \cite{Floreanini:1987as}-\cite{Witten:1996hc}
[32-37] and one could try one of these.
 For example,  adopting  the approach of \cite{Floreanini:1987as} to chiral bosons gives
 Tseytlin's formulation   \cite{Tseytlin:1990nb},\cite{Tseytlin:1990va}, while the approach of
 \cite{Siegel:1983es}  leads to the formulation of \cite{Hull:si}.
Including the coupling of the dilaton to the world-sheet
curvature, the T-duality transformation of the dilaton arises from
the transformation of the  functional measure
\cite{Tseytlin:1990nb},\cite{Tseytlin:1990va}. The resulting
theory will in general have anomalies in  both the world-sheet
conformal and Lorentz symmetries, and the vanishing of both of
these imposes \lq field equations' restricting the background
geometry. (For example, in certain cases this would lead to
separate beta-function equations for the conformal field theory of
the left-movers  and for that of the right-movers \cite{Hull:si}.)
An additional constraint follows from requiring modular
invariance. Although there are problems with writing down the
partition function for  a single chiral boson
\cite{Witten:1996hc}, the full theory here is required to have a
well-defined modular invariant partition function.

An alternative approach is through gauged sigma-models.
The world-sheet current $j_\a ^I$ defined by
 \begin{equation}
 j^I=\cp^I - L^{IJ} *(\cm _{JK}\cp^K +\cj _J)
\label{abc}  \end{equation} is conserved classically and instead
of  constraining  it to   vanish, one can gauge the corresponding
$U(1)^n$ symmetry through coupling to   world-sheet gauge fields
$A_{\a I}$, giving a formulation closely related to that of
\cite{rocek}. Again, the T-duality transformation of the dilaton
follows from the transformation of the functional measure.

It is straightforward to supersymmetrise the doubled formulation
presented here. This will be discussed elsewhere, where the
conditions for extended world-sheet supersymmetry  lead to complex
geometries similar to those of
\cite{Gates:nk},\cite{Abou-Zeid:1999em} and related to the
generalised complex structures of \cite{Hitchin}.

Here spaces with T-duality transition functions were considered
but this clearly generalises to transition functions involving
other dualities. For  M-theory  in a background  with a $T^n$
fibration  there is a U-duality symmetry  \cite{Hull:1994ys} and
so one should allow spaces with transition functions in the
U-duality group $E_n(\Z)$. Such spaces arise from U-duality
transforms of geometric backgrounds with fluxes or of T-folds, or
from reductions with U-duality twists. There are clearly relations
to the F-theory approach of
\cite{Vafa:1996xn},\cite{Kumar:1996zx}. For backgrounds which have
a Calabi-Yau fibration, one could allow transition functions that
are mirror transforms, to give mirror-folds. Indeed, when the
mirror symmetry can be understood as a T-duality, these would be
examples of T-folds. One could have patches in which there are
different string theories provided that the  transition functions
involve the dualities  relating the different string theories. For
example one could have IIA string theory on a patch $U_\a\times
S^1$
  and IIB string theory on an overlapping patch $U_\b\times S^1$ provided the transition
function is a T-duality on the $S^1$, or IIA string theory in a
neighbourhood  with a K3 fibration patched onto a heterotic string
theory in a a neighbourhood with a $T^4$ fibration.

Consider  M-theory backgrounds with a $T^n$ fibration.
These   correspond to IIA backgrounds with a $T^{n-1}$ fibration, so that
the doubled formulation   should involve
the coordinates $X^i$ of the $T^{n-1}$ and the coordinates $\ti X_i$ conjugate to the string
winding charge $w^i$  on the $T^{n-1}$.
In the M-theory picture, the string winding modes become
membrane wrapping modes and the string charge $w^i$ becomes a membrane charge
 $Z^{i \, 11}$ for a membrane wrapping the $i$'th direction on the $T^{n-1}$ and the
  M-theory circle with coordinate $X^{11}$.
  This suggests that the doubled torus of string theory  should generalise to a
  space with coordinates
  $X^i$ on the $T^n$ and coordinates $\ti X_{ij}=-\ti X_{ji}$  conjugate to the membrane wrapping charge $Z^{ij}$.
  For $n\ge 5$,   further coordinates  will be needed corresponding to the other M-theory  brane charges, e.g. coordinates
  $X_{ijklm} $ conjugate to 5-brane wrapping modes.
 For string theory, there is an explicit representation of the theory with the enlarged structure
  as a world-sheet theory allowing a concrete formulation, but for the M-theory generalisations, the absence of an  analogue of the world-sheet  formulation of the theory means that the discussion in that case has to be done in the context
 of an effective target space theory. 
    
  For example, for $n=4$ the string theory doubled torus $T^{6}$ generalises to a torus
  $T^{10}$
   with four coordinates consisting of the $T^4$ coordinates
  $X^i$ and six coordinates $\ti X_{ij}$ conjugate to membrane wrapping modes on $T^4$. The
  $O(3,3;\Z)$ or $  SL(4,\Z)$ string theory T-duality symmetry
  is enlarged to the $SL(5,\Z)$ U-duality symmetry.
  The physical space is a $T^4$ slice of the $T^{10}$, and U-duality changes the physical subspace.
  The U-fold is represented as a $T^{10}$ bundle over a 7-dimensional base space $N$, with
  $SL(5,\Z)$ transition functions, with the coordinates transforming
  according to the 10-dimensional representation of $SL(5)$.
  Space time emerges from choosing a  $T^4$ subspace of each  $T^{10}$
  fibre, and for a geometric space with transition functions in a $GL(4,\Z)$
  subgroup of $SL(5,\Z)$, this corresponds to a space time which is a $T^4$ bundle over $N$.
  For more general U-folds with non-geometric transition functions, there is a local slice
  $U_\a \times T^{4}$
  of each  patch $U_\a \times T^{10}$, but these do not fit together to form a manifold
and there is no $T^4$ sub-bundle of the $T^{10}$ bundle.

This is to be compared with the $F'$-theory constructions of \cite{Kumar:1996zx}, which involve a $T^5$ bundle over $N$. The $SL(5,\Z)$ transition functions can be used to construct either a $T^5$ bundle or a $T^{10}$ bundle, but in the   $T^{10}$ construction discussed here, the 
extra dimensions play a physical role and are related to   brane charges.
In particular, non-trivial U-duality transition functions mean that there is no clear distinction between physical spacetime coordinates and auxiliary coordinates, and there can be 
configurations where each  of the ten  torus coordinates becomes a physical spacetime coordinate somewehere, but only four of the ten are simultaneously physical in any given local patch.
Similar remarks apply to all the examples discussed here.

 As a further example, consider M-theory on $T^7$.
 Viewed as  IIA string theory on $T^6$, there is a doubled torus $T^{12}$ and for M-theory this
  is enlarged to a $T^{56}$ with 7 coordinates
 $X^i$ on $T^7$, 21 coordinates $X_{ij}$ conjugate to the M2-brane charge,
 21 coordinates $X_{i_1...i_5}$ conjugate to the M5-brane charge, and
 7 coordinates $X_{i_1...i_6}$ conjugate to the Kaluza-Klein monopole charge \cite{Hull:1997kt}.
 The enlarged spacetime would be a $T^{56}$ bundle over a four dimensional base $N$
 with transition funtions in $E_7(\Z)$.
 Physical space consists of choosing a local $T^7$ within each $T^{56}$ fibre, and
 these will not fit together to form a $T^7$ bundle over spacetime in general unless the
 structure group is in the geometric group $GL(7,\Z)$ acting on the $X^i$.

 In \cite{Hull:2003mf},   it was suggested that  spacetime can be pictured as arising as a surface
 in a larger space with a coordinate conjugate to each of the BPS brane charges in the superalgebra,
 with duality acting to change the embedding of the  spacetime slice in the bigger space.
 The construction given here gives a concrete realisation of this in the context of perturbative
 string theory, and it seems to give the natural framework for discussing non-geometric backgrounds.

The discussion here has been concerned mainly with the case of spaces with a
torus fibration, but can be generalised  to those with a fibration
by Calabi-Yau spaces or other spaces on which a stringy duality
acts. In the case of a torus, the duality naturally leads to the
doubling of the internal space in string theory, giving the
picture of space-time as a surface in a larger space. It would be
interesting to see to what extent this could be extended to more
general spaces which do not have a torus or Calabi-Yau fibration
and ask whether a doubled geometry, or the extension arising in
M-theory, has a role to play in that case too. In
\cite{Dabholkarx}, the possibility of non-geometric string
backgrounds which do not have torus fibrations will be explored.

\end{document}